# LEARNING CURVES
# IN NATURAL AND ARTIFICIAL SYSTEMS

**Novikov D. A.**[1]
*(V. A. Trapeznikov Institute of Control Sciences of the Russian Academy of Sciences, Moscow)*

*Abstract: A general mathematical model of experience formation is considered, in which the learning level is defined as the probability that the next value of the uncertainty factor has already been mastered and not forgotten. Within this model the shape of the learning curve is derived, and it turns out to be determined primarily by the distribution of situation occurrence: a uniform distribution yields an exponential curve, whereas a power-law distribution yields a power-law curve whose exponent equals the ratio of the distribution exponent to that same exponent increased by one. Both mechanisms generating power-law curves, established earlier in the analysis of debugging, assembly and repair activities, are shown to have been independently rediscovered in deep learning theory as the "frequency" and "spectral" mechanisms of scaling laws for artificial neural networks, while the exponential curves of machine learning (contraction operators, strongly convex optimization, stationary reinforcement) are generated by the same constancy of the contraction coefficient that was revealed half a century earlier in models of radioactive decay, the first-order lag element, and others. A solution is given for the problem of choosing a training distribution that differs from the deployment distribution. A methodological consequence is discussed: three distinct mechanisms generate curves indistinguishable in shape, so an observed learning curve does not allow one to judge the "internal" structure of the learning system.*

## 1. Introduction

The *learning curve* – the dependence of some quality characteristic of an action on the number of its repetitions – has been discovered

---

[1] *Dmitry Novikov, Doctor of Technical Sciences, Professor, Academician of the Russian Academy of Sciences, Director of the Institute of Control Sciences RAS (novikov@ipu.ru).*

independently in several non-overlapping subject areas, under a different name each time.

In 1885 H. Ebbinghaus obtained the first quantitative description of the forgetting of memorized material [24]. In 1899 W. Bryan and N. Harter measured the dynamics of mastering the telegraphic code and discovered the "*plateau*" – a stretch on which the external indicators do not grow while the skill is being internally restructured, moving from letters to words and from words to phrases [17]. In 1919 L. Thurstone proposed a *hyperbolic* approximation of the accumulated error [70], and in 1943 C. Hull an *exponential* dependence of habit strength on the number of reinforcements [32]. In 1936 the engineer T. Wright noticed that with every doubling of the cumulative output of aircraft, the labor input per unit drops by the same fraction [80]; the psychological curve thereby became an economic law and a tool for strategic cost planning. In the 1950s R. Bush and F. Mosteller described the behavior of an animal (a rat) in a maze by a difference equation for the probability of a correct response [18].

In the 1950s–1960s, learning curves appeared in artificial systems. M. Tsetlin studied the behavior of finite automata in random environments [71]; W. Ashby built the homeostat [3, 4]; F. Rosenblatt the perceptron [58]; Ya. Tsypkin created a general theory of learning systems in the language of stochastic approximation [73, 74]. The curves of machines proved qualitatively indistinguishable from those of the animal and the telegraphist, which ruled out peculiarities of physiology as the explanation.

Finally, over the last decade the training of artificial neural networks (ANNs) has come to be described by *scaling laws*: it has been established empirically [30, 34] that the error decreases as a power law in the data volume, the number of parameters, and the compute budget, with exponents stable over five to six orders of magnitude of resource variation. An engineer planning the training of the next large language model is using Wright's curve, with tokens in place of aircraft.

What an animal in a maze, a telegraphist, an aircraft factory, and a transformer have in common is not explained by the answer "they all learn": they learn differently and on entirely different material. Over a century, several dozen mathematical models have accumulated, and almost every one explains its own fragment by its own means: one derives the curve from the kinetics of a monomolecular reaction, another from the limited capacity of a communication channel, a third from the law of radioactive decay, a fourth from game theory, and a fifth simply postulates an exponential (see the survey in [49]). Each is convincing on its own, but together they amount to a collection with no theory behind it. As early as 1935, E. Guthrie wrote that no general learning curve exists and that everything depends on the learner's prior experience [28].

The present paper starts from the opposite thesis. It is claimed that the commonality of the regularities of iterative learning in natural and artificial systems is formal rather than metaphorical: the models listed above are *different aspects ("sections") of one construction*, corresponding to different assumptions about the environment and the learner. The language in which this claim can be stated precisely is that of experience rather than of the curve.

In substance the construction is as follows. The learner – an individual, a collective, an automaton, a neural network – encounters, time after time, situations drawn from some set. Experience is the set of situations for which the learner already knows what to do. The learning level is the probability that the next realized situation turns out to be among those already mastered. Nothing more needs to be postulated: the learning curve is derived from these definitions, and its shape is determined by how often the different situations occur and how reliably they are mastered and how they are forgotten.

Hence the main result of this work, polemical with respect to the prevailing practice: **the shape of the learning curve characterizes chiefly the environment – more precisely, the occurrence distribution of the situations the learner encounters – and only secondarily the learner.** An exponential corresponds to a homogeneous environment; a power-law curve to an environment with a "heavy tail" of rare situations, that is, to the complex character of the activity being mastered. A century of the psychology of learning has been spent on attempts to extract the learner's internal structure from the shape of the curve – from the strengthening of stimulus–response bonds to chunking; the claim that the curve's shape primarily reflects the statistics of the task rather than the architecture of the solver is empirically testable and, if true, devalues a substantial part of these attempts. H. Simon already observed that the apparent complexity of a system's behavior over time mostly reflects the complexity of the environment in which the system operates [65].

The structure of the paper follows this logic. Section 2 introduces the descriptive language: iterative learning, mismatch, the exponential curve, and the rate of learning as the invariant of an equivalence class. Section 3 presents the general model of experience formation and its simplest special cases. Section 4 separates the three distinct “sections” that in machine learning go under the single term “learning curve”. Sections 5 and 6 study the first two sections (by sample size and by number of iterations, respectively) and establish under what conditions the curve is a power law and under what conditions it is exponential. Section 7 is devoted to scaling laws and the problem of optimally allocating the compute budget; Section 8 to reinforcement learning; Section 9 to nonmonotone curves that systematically violate W. Ashby's principle of “never losing what has been gained”. Section 10 gives a normative answer to the question of partitioning the situation space and of typical solutions. Section 11 poses and solves the problem of choosing a training distribution that differs from the deployment one. Section 12 contains an explicit correspondence between the basic concepts of the general model of activity and those of machine learning. Sections 4–12 do not assume the reader is familiar with machine learning: the general problem statement of that discipline and its vocabulary are given at the beginning of Section 4, and specialized terms are defined in footnotes at their first use.

The author is grateful to M.V. Belov and A.V. Makarenko for valuable comments.

## *2. The Descriptive Language: Iterative Learning and Learning Curves*

*Iterative learning* (IL) is the kind of learning based on the repeated performance, by the learning system, of the same actions (trials) under constant external conditions and an unchanging goal [49]. Both constancy conditions are essential for methodological reasons. Quantitative relations between a system's input and output can be established only when the change in the response is caused by a change in a single input signal; if the information about the result of the previous trial and the parameters of the environment changed simultaneously, learning outcomes at different moments of time would be incomparable – one could not tell whether a given output value was realized because the learner had changed or because the conditions had.

Two aspects of learning are distinguished: the *outcome* aspect (achieving the required quality of performance at acceptable costs of time and energy) and the *process* aspect (adaptation, adjustment of the learning system). Below we deal exclusively with outcome characteristics; adaptation characteristics often have quite different dynamics [48].

As the outcome characteristic we take a *learning-level criterion*, whose role in the training of real systems is played by quantities of four groups [48]: temporal (time to perform an action, reaction time, error-correction time); rate-type (labor productivity, reaction speed – quantities inverse to time); accuracy-type (error magnitude in physical units, number of errors, error probability); and informational (amount of material memorized, amount of information processed). In all these cases we have either an error function or a "learnedness" characteristic reducible to an error function, so for generality of exposition the quantity of interest will be called the *mismatch*. Thus, the time to perform an action is interpreted as a mismatch if the latter is understood as the difference between the current and the minimum possible values.

Iterative learning is typically characterized by decelerating asymptotic curves approximated by *exponential dependences* of the form

(1) $x(t) = x_\infty + (x_0 - x_\infty)\, e^{-\gamma t}, \qquad t \geq 0,$

or by sequences $x_n = x_\infty + (x_0 - x_\infty)e^{-\gamma n}$, $n = 0, 1, 2, \ldots$, where $t$ is the learning time, $n$ is the number of trials, $x_0$ is the initial value of the mismatch, $x_\infty$ is its asymptotic value (in biological systems this quantity is interpreted as the *physiological limit of learning*), and $\gamma \geq 0$ is the *rate of learning*, whose dimension is inverse to time or to the number of iterations.

Depending on the relation between $x_0$ and $x_\infty$, expression (1) describes both decreasing and increasing curves. The quantitative characteristics $x_0$, $x_\infty$, $\gamma$ depend on many factors – the complexity and properties of the learning system, the properties of the environment, the training methodology employed. We are interested in the qualitative shape of the curve, so in what follows we use the normalized dependences

(2) $x(t) = e^{-\gamma t}, \qquad L(t) = 1 - e^{-\gamma t},$

obtained from (1) by a linear transformation. If $x(t)$ is the mismatch, then $L(t) = 1 - x(t)$ is the *learning level*. Thus, when speaking of a *learning curve* (LC), we mean a family of curves equivalent up to a linear transformation; the invariant of the equivalence class is the rate of learning $\gamma$.

Exponential curves are the most common, though other kinds occur as well. L. Thurstone, on animal-learning data, proposed a hyperbolic approximation $x(n) = \alpha n/(b + n)$ for the accumulated error [70]; the averaged curves obtained within stimulus-sampling theory are close to an exponential-type function [5]. The question of which of the two approximations – exponential or power-law – better describes empirical data has been debated for several decades [29, 47] and, as will be shown below, has no answer independent of the structure of the environment.

Other kinds of LC exist as well; for example, the analogy with an autocatalytic reaction [49] leads to the *logistic curve*

(3) $x(t) = \frac{x_0\, x_\infty}{x_0 + (x_\infty - x_0)\, e^{-\gamma t}},$

which, depending on the relation between the initial and final values of the mismatch, can be either increasing or decreasing.

Finally, the mastering of complex activity is almost always approximated by *power-law* curves [47, 80]. The three types of curves, with comparable parameters, are shown in Fig. 1.

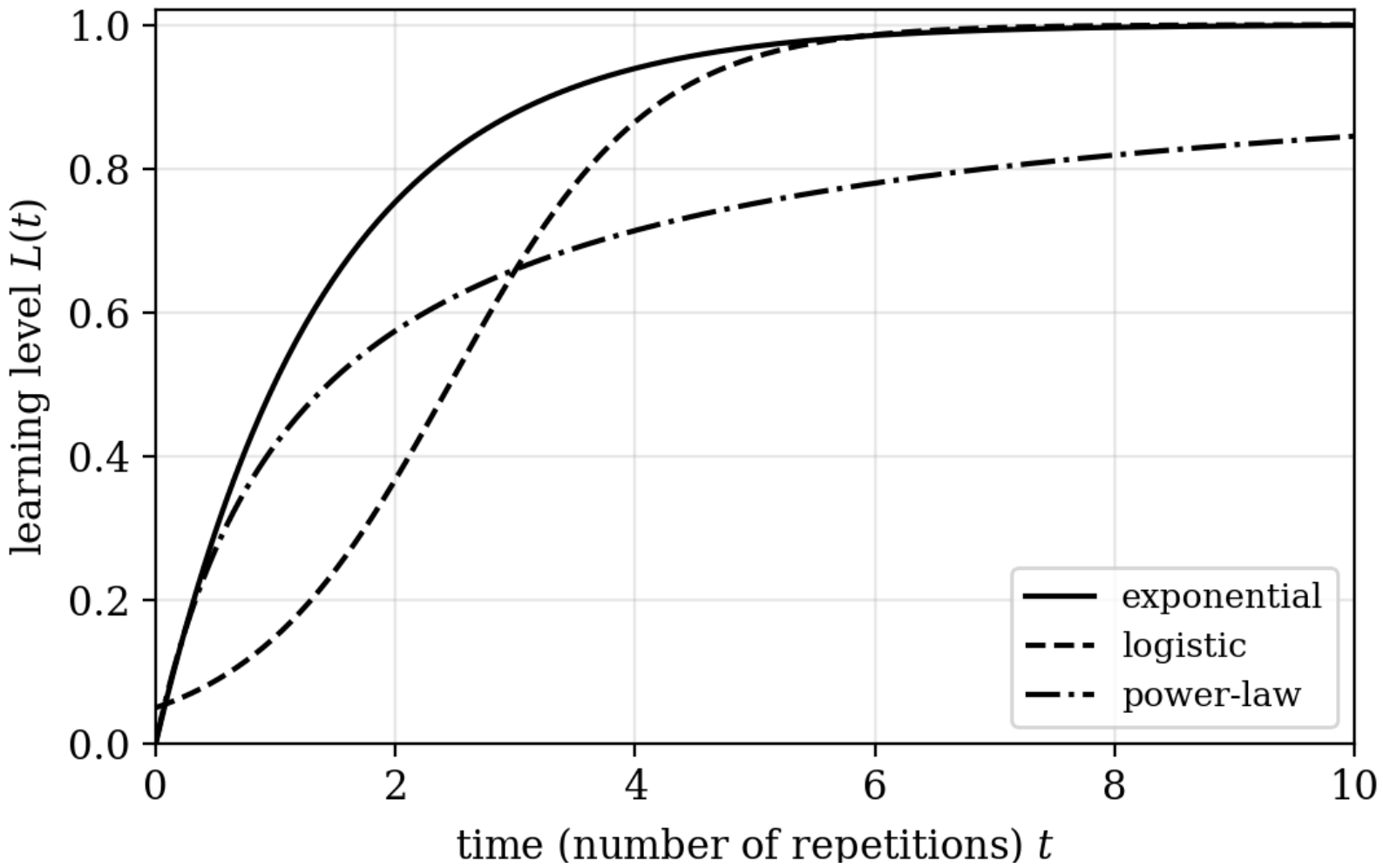

*Fig. 1. Three types of normalized learning curves*

Two further circumstances matter for what follows.

First, for comparatively complex kinds of activity the curve may have an intermediate *plateau*, whose presence is explained by the learning system's hidden search for new ways of performing actions and its preparation for the transition to a qualitatively new strategy [17]. Formally, such a curve is a composition of two successive concave increasing curves, or an exponential curve later replaced by a logistic one. A plateau can also arise as an artifact of averaging different curves over learners or over skills.

Second, some of the observed features are generated by the choice of measurement scale and have nothing to do with the learning mechanism. When dichotomous scales are used (a critical error level is set arbitrarily and the action is deemed either correct or incorrect), or when one passes to inverse quantities (speed as the reciprocal of time), the exponential curve turns into a logistic one [48]. The appearance of logistic curves in this case is unnatural and is eliminated by choosing an appropriate scale and units of measurement. We shall need this remark in Section 9 when discussing "emergent" effects in large language models.

## *3. The General Model of Experience Formation*

The models listed in Sections 1 and 2 share a common methodological defect: decelerating asymptotic behavior is either postulated in them or built into the linearity assumptions on the right-hand side of the corresponding differential (difference) equation. This defect is removed by the model of experience formation [11, 12], briefly presented below.

Let an *active system* (AS) consist of a set of $N = \{1, \dots, n\}$ *active elements* (AEs). In each period, every AE observes one of $K$ possible values of the *uncertainty factor* (UF) – a formalization of the "situation" in which action is required. The state of the AS is described by a matrix of $\mathbf{v}(t) = \| v_{ik}(t) \|$ binary variables: $v_{ik}(t) = 1$ if, after the $t$-th period, the $i$-th AE has mastered the technology (formed experience) for the $k$-th value of the UF.

The evolution mechanism is as follows. If the UF has taken the value $k$, then for any value $l$ not yet mastered by the $i$-th AE, *experience formation* occurs with the *probability of mastering* $w_{ikl}\{\cdot\}$, and for any mastered value, *forgetting* occurs with the *probability of forgetting* $u_{ikl}\{\cdot\}$; in general, both probabilities depend on time and on the history of the AS's states. Writing $q_{ik}(t) = \Pr\{v_{ik}(t) = 1\}$ and averaging the probabilities of mastering and forgetting over the UF values (notation $W_{ik}$, $U_{ik}$), by the total probability formula we obtain

(4) $q_{ik}(t+1) = W_{ik}\big(\mathbf{q}(t)\big) + \big[1 - W_{ik}\big(\mathbf{q}(t)\big) - U_{ik}\big(\mathbf{q}(t)\big)\big]\, q_{ik}(t)$

with known initial conditions $\mathbf{q}(0)$; in continuous time, (4) is replaced by the system of differential equations

$$\dot{q}_{ik} = W_{ik}(\mathbf{q}) - \big(W_{ik}(\mathbf{q}) + U_{ik}(\mathbf{q})\big)\, q_{ik}.$$

The following definition is central to the construction. The value of the *individual experience criterion* ("learning level") of the $i$-th AE is the probability that a UF value is realized that the AE has already encountered, successfully mastered, and not forgotten:

(5) $L_i(t) = \sum_{k=1}^{K} p_{ik}\; q_{ik}(t) = 1 - \sum_{k=1}^{K} p_{ik}\; \big(1 - q_{ik}(t)\big).$

The *collective experience* criteria are defined as the probability that the realized UF value has been mastered by at least one AE ($L_{\max}$) or by every AE ($L_{\min}$). The sequence of experience-criterion values is what is called the *experience curve*, or the *learning curve*.

Definition (5) moves the question of the learning curve's shape from the category of postulates into the category of consequences: the curve is a sequence of probabilities that the next UF value has already been mastered and not forgotten, and its form is completely determined by the distribution $\{p_k\}$ of the probabilities of these values together with the probabilities of mastering and forgetting. Consider the simplest special cases (the numbering, description and notation of models O1, etc., follow [11]).

**Model O1** (one AE, stationary parameters, probability of mastering equal to one, no forgetting). Denoting by $p_k > 0$ the probability of presenting the $k$-th UF value, from (4) we obtain $q_k(t+1) = p_k + (1-p_k)q_k(t)$, whence $1 - q_k(t) = (1-p_k)^t$ and

(6) $L(t) = 1 - \sum_{k=1}^{K} p_k \ (1-p_k)^t$.

In the special case of the uniform distribution $p_k = 1/K$, expression (6) gives

(7) $L(t) = 1 - (1-1/K)^t \approx 1 - e^{-t/K}$,

that is, **an exponential learning curve with rate** $\gamma = 1/K$ – the curve that was independently derived in a dozen twentieth-century models from the kinetics of chemical reactions, the law of radioactive decay, and the transfer function of a first-order lag element [49]. Here it follows from the homogeneity of the environment and need not be postulated.

**Model O4** (stationary probabilities of mastering $w_k$ and forgetting $u_k$). From (4) it follows that $q_k(t+1) = p_k w_k + \big(1 - p_k(w_k + u_k)\big)\, q_k(t)$, and in the homogeneous case $w_k \equiv w$, $u_k \equiv u$

(8) $L(t) = \frac{w}{w+u}\left[1 - \sum_{k=1}^{K} p_k \ \big(1 - p_k(w+u)\big)^t\right]$.

Curve (8) has a nonzero asymptote $w/(w+u) < 1$: **the "physiological limit of learning" results from a balance between mastering and forgetting rather than from a property of the learner as such.**

**Model O3** (the probability of mastering depends on the learning level attained). Let $w\big(\mathbf{q}(t)\big) = g\big(L(t)\big)$. Then, for the uniform distribution, (4) yields the difference equation $\Delta L(t) = \frac{1}{K}\, g\big(L(t-1)\big)\big(1 - L(t-1)\big)$, and depending on the form of the function $g(\cdot)$ all three classical curve types are realized (Table 1).

*Table 1. Shape of the learning curve as a function of the law governing the probability of mastering (model O3)*

| Probability of mastering $g(L)$ | Dynamics equation | Learning curve |
|---|---|---|
| $g(L) = \gamma K$ | $\Delta L = \gamma(1 - L)$ | Exponential |
| $g(L) = \mu KL$ | $\Delta L = \mu L(1 - L)$ | Logistic |
| $g(L) = \eta K(1 - L)^a$ | $\Delta L = \eta(1 - L)^{1+a}$ | Power-law |

Thus a single construction generates exponential, logistic and power-law curves as special cases corresponding to different assumptions about the environment (the distribution $\{p_k\}$) and about the learner (the laws $w(\cdot)$, $u(\cdot)$). The exposition that follows shows that these special cases, up to notation, have been obtained independently in machine learning.

The model's limitations should be stated as well. The binarity of the variables $v_{ik}$ means that the model represents neither skill transfer, nor interference, nor generalization – the phenomena that the psychology of learning regards as decisive for skill formation, and that in machine learning make up an essential part of the discipline's subject matter. Abandoning binarity and introducing a more complex, possibly hierarchical, structure on the set of UF values appears to be a promising direction for further research. This can (and should) be done using two- and more-level models of *learning to learn* – see [11, 10].

## 4. The General Model of Machine Learning and the Three Arguments of the Learning Curve

This section begins with the general problem statement of machine learning. It is given in a form directly comparable with the general model of experience formation of Section 3 and fixes a unified system of notation used in Sections 4–12. A reader familiar with machine learning should pay attention primarily to the renamings: they are collected in Table 2 and are caused by the fact that a number of symbols conventional in that field are occupied above for other concepts.

*Machine learning* from precedents is the collection of methods for constructing algorithms that are tuned (“trained”) on a set of examples instead of being designed in advance for each specific situation. The general model of machine learning is defined by the following objects.

– The *set of situations* (precedents) Ω. An element ω ∈ Ω is a *situation* (in machine learning – an *instance*, or *input*) presented to the learning system and described, as a rule, by a set of *features*[2]. In the terms of Section 3, the set Ω is the set of values of the uncertainty factor (UF).

– The *set of outcomes Z*. An element $z \in Z$ (in machine learning – a *label*, or *ground truth*) is what the system should have produced in situation ω.

– The *set of actions Y*. An element $y \in Y$ is the system's answer (action); in most problems $Y = Z$.

– The *distribution P* on $\Omega \times Z$. It is unknown; only examples are available – this is the uncertainty that learning overcomes.

– The *set of admissible technologies T*. An element $\tau \in T$, $\tau: \Omega \to Y$ (in machine learning – an *algorithm*, *hypothesis*, or *model*), is a rule assigning an action to every situation. As a rule, $T$ is parameterized: $\tau = \tau(\cdot, \theta)$, where θ is the *parameter vector* and $M$ is the number of these parameters, called the *model size*[3].

– The *loss function* $\ell(y, z) \geq 0$ – the amount of damage incurred when answer $y$ is produced while outcome $z$ was required. Typical examples: $\ell(y, z) = (y - z)^2$ in prediction problems and $\ell(y, z) = 1\{y \neq z\}$ in recognition problems.

The average loss of technology τ is called, in machine learning, the *risk*:

(9) $F(\tau) = E[\ell(\tau(\omega), z)], \; F^* = inf_\tau F(\tau), \; F_T = inf_{\tau \in T} F(\tau).$

Here $F^*$ is the smallest risk attainable at all (in probability theory it is called the *Bayes risk*), and $F_T$ is the smallest risk attainable within a given class of technologies $T$. A *training set* is a collection $S = \{(\omega_1,$ $z_1$), …, ($\omega_D$, $z_D$)} of $D$ independent realizations of the distribution $P$; the number $D$ is called the *sample size*. A *structural learning algorithm* $\mathcal{A}$ is a rule assigning a technology to a sample: $\tau_D = \mathcal{A}(S)$.

---

[2] *A feature is a measurable characteristic of a situation, expressed as a rule by a number, for example: the brightness of an image pixel, a sensor reading, the presence of a given word in a text. An ordered set of features (a vector) is identified with the situation itself, and the set Ω with a subset of a finite-dimensional vector space.*

[3] *The class of technologies most widespread at present is artificial neural networks: superpositions of linear transformations and nonlinear functions of one variable. The number of hidden layers of such a superposition is called its depth (hence deep learning), the way they are connected its architecture, and the collection of coefficients of the linear transformations its weights, that is, the parameter vector θ. Present-day values of M range from $10^6$ to $10^{12}$.*

Since the distribution $P$ is unknown, the risk (9) cannot be computed directly. Instead, one computes the average loss on the sample itself – the *empirical risk* – and the technology is chosen so as to minimize it:

(10) $\tau_D \in \operatorname{Arg\,min}_{\tau \in T} F_S(\tau)$, где $F_S(\tau) = \frac{1}{D}\sum_{i=1}^{D} \ell(\tau(\omega_i), z_i)$.

Condition (10) is called the *empirical risk minimization* principle. It constitutes the formal content of the word "learning": the system chooses the technology (first the model class, then the parameters of a specific model) that is best on the examples presented to it, in the expectation that this same technology will also prove good in a case not yet presented[4].

**The learning problem.** Given: a class of admissible technologies $T$, a loss function $\ell$, and a sample $S$ of size $D$ generated by an unknown distribution $P$. Required: construct a technology $\tau_D \in T$ minimizing the risk (9). This problem cannot be solved exactly, and the measure of what remains unattained is the *excess risk*, which admits the decomposition

(11) $F(\tau_D) - F^* = (F_T - F^*) + (F(\tau_D) - F_T)$,

in which the first term – the *approximation error* – reflects the poverty of the class of available technologies and does not depend on the sample size, while the second – the *estimation error* – reflects the incompleteness of experience and decreases as $D$ grows. The theory of learning curves is concerned with the second term. In the terms of Section 2, the approximation error is the learning limit $x_\infty$ of expression (1), and the estimation error is the decreasing part of the mismatch.

The scheme of the general model is shown in Fig. 2. The environment generates "situation – required outcome" pairs; a finite set of such pairs forms the training set; the learning algorithm tunes the technology's parameters on it; the technology, receiving a new situation, produces an action; the loss function compares the action with the required outcome, and the result of this comparison is used for further tuning of the parameters. The loop closing through the loss function onto the learning algorithm is the adaptive control loop known from the cybernetic models of the mid-twentieth century, with the technology itself serving as the "controlled object".

---

[4] *The gap between the empirical risk and the risk is called overfitting: the technology memorizes the presented examples instead of uncovering the regularity. To detect it, the sample is split into a training set and a test (hold-out) set; the risk is estimated on the test set, which took no part in tuning the parameters. A technology's ability to keep its loss small on examples never presented to it is called its generalization ability.*

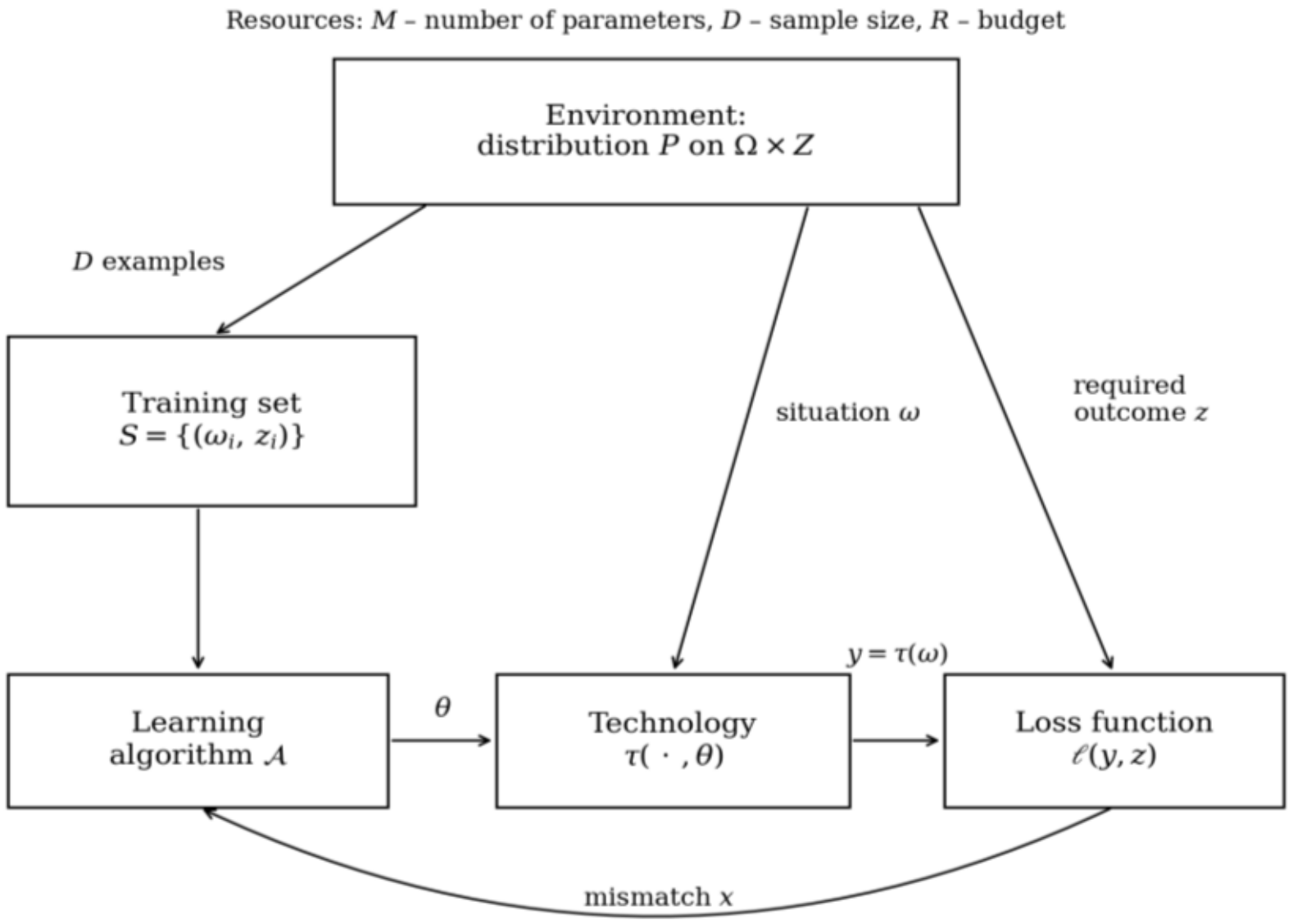

*Fig. 2. General scheme of machine learning*

The three main special cases of the general model differ in what is available to the learning system. In *supervised learning*, the sample contains pairs (ω, *z*), that is, the required outcome is presented explicitly. In *unsupervised learning*, required outcomes are absent and the structure of the distribution *P* itself is recovered. In *reinforcement learning*, instead of the required outcome the system receives only a scalar assessment of its action – the *reinforcement* (see Section 8), and this assessment may arrive with a delay. The first case corresponds directly to the experience-formation model of Section 3, the third to the classical models of animal learning.

**Three parameters.** We shall assume that the attainable quality of learning is determined mainly by three parameters ("resources"): the *model size M* (the number of its parameters), the *sample size D* (the number of examples), and the *compute budget R* – the number of elementary operations spent on training. The relation between these three resources is the subject of Section 7, and the three "sections" of the learning curve distinguished below correspond to them.

The notation used in Sections 4–12 is collected in Table 2. Throughout, we keep to the notation system of Sections 2 and 3; the third column gives the notation adopted in the machine learning literature where it differs from ours.

*Table 2. Notation for Sections 4–12*

| Notation | Meaning | In machine learning |
|---|---|---|
| $\omega \in \Omega$ | situation (input, instance); UF value | $x \in X$ |
| $z \in Z$ | required outcome (label, ground truth) | $y$ |
| $y = \tau(\omega)$ | action (answer) of the learning system | $\hat{y}$ |
| $\tau \in T$ | technology of activity (hypothesis, model) | $f \in F$, $h \in H$ |
| $\theta$ | parameter vector of the technology | $\theta$, $w$ |
| $M$ | model size – number of parameters | $N$ |
| $D$ | training sample size – number of examples | $n$, $N$, $D$ |
| $S$ | training set | $S$, $D$ |
| $R$ | compute budget | $C$ |
| $\ell(y, z)$ | loss function on a single example | $\ell$ |
| $F(\tau)$ | risk – average loss | $R(f)$, $L$ |
| $F_S(\tau)$ | empirical risk – average loss on the sample | $\hat{R}(f)$ |
| $F^*$, $F_T$ | Bayes risk; minimum risk over the class $T$ | $R^*$ |
| $x(D)$, $x_t$ | mismatch (excess risk) | $\varepsilon(n)$ |
| $x_\infty$ | learning limit – irreducible loss | $E$ |
| $L$ | learning level, $L = 1 - x / x(0)$ | 1 – error |

| Notation | Meaning | In machine learning |
|---|---|---|
| $t$ | iteration number of the algorithm | $t$; epoch |
| $\eta$, $\eta_t$ | step size and its schedule | $\eta$ |
| $\mu$, $\Lambda$ | strong convexity and smoothness parameters | $\mu$, $L$ |
| $G$, $\lambda_i$ | Hessian and its spectrum | $H$, $\lambda_i$ |
| $\sigma^2$ | variance of the gradient-estimation noise | $\sigma^2$ |
| $\alpha$ | exponent of the power-law learning curve | $\alpha$ |
| $\beta$ | exponent of the situation-occurrence distribution | $\alpha$ |
| $\zeta$ | normalizing constant of the Zipf distribution | – |
| $K$ | number of distinguished UF values (partition cells) | $K$ |
| $p_k$, $\pi_k$ | deployment and training distributions | $p$, $q$ |
| $V_t$, $r_t$ | value estimate and reinforcement | $V$, $r$ |
| $\chi$ | contraction coefficient of the Bellman operator | $\gamma$ |
| $l$ | Lipschitz constant | $L$ |
| $\varphi(y, \omega)$ | efficiency of decision $y$ in state of nature $\omega$ | – |

One divergence of notation deserves separate mention. The quantity denoted in machine learning by the letter $L$ (loss) is the *loss*, that is, the error; above, the letter $L$ denotes the *learning level*, the quantity complementary to the error. To avoid confusion, loss is everywhere below denoted by the letter $F$ – the same as the expected error in (33) – while the letter $L$ is reserved for the learning level; the relation between them is given by the equality

$$L(D) = 1 - x(D) / x(0).$$

In Sections 2 and 3 the learning curve is a function of one argument – time or the number of repetitions. In machine learning the term *learning curve* [53] covers three distinct sections of a single phenomenon, which must be strictly distinguished [43, 78].

Let $S_D$ be a training set of size $D$ generated by the distribution $P$; $F(\tau) = \mathbb{E}_P\, \ell(\tau(\omega), z)$ the risk; $F^* = \inf_\tau F(\tau)$ the smallest attainable risk; $\tau_D$ the model trained on the sample $S_D$. The *sample-wise learning curve* is the function

(12) $x(D) = \mathbb{E}\, F(\tau_D) - F^*.$

The quantity $x(D)$ is the *excess risk*; in the notation of Section 2 it is the *mismatch*, and the normalized *learning level* equals $L(D) = 1 - x(D)/x(0)$. Empirically (see the survey in [78]), in the overwhelming majority of problems

(13) $x(D) \approx x_\infty + A\, D^{-\alpha}, \qquad \alpha > 0,$

that is, the learning curve is a power law rather than an exponential – the situation that below is linked to the complex character of the activity being mastered.

The second section is the *optimization curve* $x(t)$, where $t$ is the iteration number of the algorithm for a fixed sample. The third is the *scaling law*

(14) $F = F(M, D, R),$

where $M$ is the number of model parameters, $D$ the volume of training data, $R$ the compute budget.

The first section answers the question "how much experience is needed", the second "how many repetitions are needed", the third "how much resource is needed". In the terms of Section 2, the first two are outcome characteristics of learning; the third is a *control problem*. Mixing these sections up is a common source of terminological confusion: statements about the "exponential" or "power-law" character of a learning curve are meaningful only with the argument explicitly specified.

## *5. Sample-Wise Learning Curves*

**Problem statement.** Given: a class of admissible technologies *T*, a loss function $\ell$, an unknown distribution *P*, and a learning algorithm $\mathcal{A}$. Required: find the form of the dependence of the excess risk *x*(*D*) on the training sample size *D*, specify the conditions under which this dependence is a power law and under which it is exponential, and express the exponent α through characteristics of the class *T* and the distribution $P$[5]. In practical terms, the problem is how many examples must be collected to reach a given accuracy, and/or by what factor the sample should be enlarged to reduce the error a given number of times, say by half.

Classical statistical learning theory [75, 76] gives bounds uniform over the hypothesis class. If the class has *Vapnik–Chervonenkis dimension* $d_{VC}$, then with probability at least $1 - \delta$

(15) $x(D) \leq c\sqrt{\frac{d_{VC}+\ln(1/\delta)}{D}}$,

that is, the learning curve is a power law with exponent $\alpha = 1/2$. In the realizable case, when the class contains a hypothesis with zero error, the rate doubles:

(16) $x(D) \leq c\,\frac{d_{VC}\ln D}{D}, \qquad \alpha = 1.$

Intermediate values of $\alpha \in [1/2;\ 1]$ are obtained under A.B. Tsybakov's noise condition with parameter $\kappa \geq 1$ [72]:

(17) $x(D) \asymp D^{-\kappa/(2\kappa-1)}$.

In the nonparametric setting, when the sought dependence belongs to a smoothness class $s$ in a space of dimension $d$, the minimax rate equals

(18) $x(D) \asymp D^{-2s/(2s+d)}$,

whence $\alpha = 2s/(2s+d) \to 0$ as $d \to \infty$ – the *curse of dimensionality*. This effect partly explains why learning curves change slowly in high-dimensional problems: the exponent is small, and doubling the sample yields little gain.

Deep ANNs sidestep the curse of dimensionality thanks to the compositional structure of the functions being approximated [82] – a remark that corresponds directly to the proposition of A.N. Leontiev's psychological theory of activity that an action is composed of operations [39] (see also Section 7).

*Exponential* learning curves do occur in machine learning as well, but only under strong additional assumptions: class separability with a nonzero margin, or active (non-random) selection of training examples, when each trial is guaranteed to shrink the region of uncertainty by a fixed factor, $x(D) \le c\,e^{-\gamma D}$. The same dualism, discussed below, appears here: the exponential corresponds **to the removal of a fixed share of uncertainty per trial, the power law to the exhaustion of the "heavy tail" of rare situations.** In the statistical mechanics of learning [1, 61], four qualitatively different types of learning curves were obtained for the perceptron, including curves with a first-order *phase transition* – a formal analogue of the "plateau" of Section 2.

---

[5] *Here and below the standard asymptotic notation is used: g = O(h) means that the ratio g/h is bounded from above; g $\asymp$ h that it is bounded from above and below by positive constants; g $\simeq$ h denotes asymptotic equality; Θ(h) the exact order of growth. A tilde over a symbol (Õ) means that logarithmic factors are also omitted.*

The main result of this section links the exponent of the power-law curve to the structure of the environment.

**Proposition 1.** *Let the UF values* $k = 1, 2, \dots$ *be presented independently with probabilities* $p_k$*, and let the corresponding component of experience be deemed mastered if the situation has occurred at least once in D presentations (model O1 of Section 3). Then the unmastered fraction equals*

(19) $x(D) = \sum_{k\ge 1} p_k\,(1-p_k)^D \approx \sum_{k\ge 1} p_k\, e^{-p_k D}$,

*and the following holds:*

*– for the uniform distribution* $p_k = 1/K$ *the learning curve is exponential,* $x(D) = e^{-D/K}$*;*

*– for the power-law (Zipf) distribution* $p_k = \zeta\, k^{-(1+\beta)}$, $\beta > 0$, *the learning curve is a power law,*

(20) $x(D) \simeq \frac{\zeta^{1/(1+\beta)}}{1+\beta}\,\Gamma\left(\frac{\beta}{1+\beta}\right) D^{-\beta/(1+\beta)}$,

*that is, the exponent equals* $\alpha = \beta/(1+\beta)$.

**Proof.** Replacing the sum in (19) by an integral and setting $a = 1+\beta$, we have $x(D) \approx \int_1^{\infty} \zeta\, y^{-a}\, e^{-\zeta D y^{-a}}\, dy$. The substitution $u = \zeta D y^{-a}$, under which $y = (\zeta D/u)^{1/a}$ and $dy = -\frac{1}{a}(\zeta D)^{1/a} u^{-1/a-1} du$, gives $x(D) \approx \frac{(\zeta D)^{1/a}}{a\,D}\int_0^{\infty} u^{-1/a}\, e^{-u}\, du = \frac{\zeta^{1/a}}{a}\,\Gamma\left(1-\frac{1}{a}\right) D^{1/a-1}$. Since $1 - 1/a = \beta/(1+\beta)$, we obtain (20). •

Proposition 1 demonstrates the paper's main thesis: one and the same model **of the learner, under different environment statistics, generates qualitatively different learning curves.** An illustration is given in Fig. 3: in double logarithmic coordinates the curve for the uniform distribution "breaks off" (an exponential), whereas the curves for Zipf distributions settle onto straight lines with slopes $-1/3$ and $-1/2$, corresponding to $\beta = 0{,}5$ and $\beta = 1{,}0$.

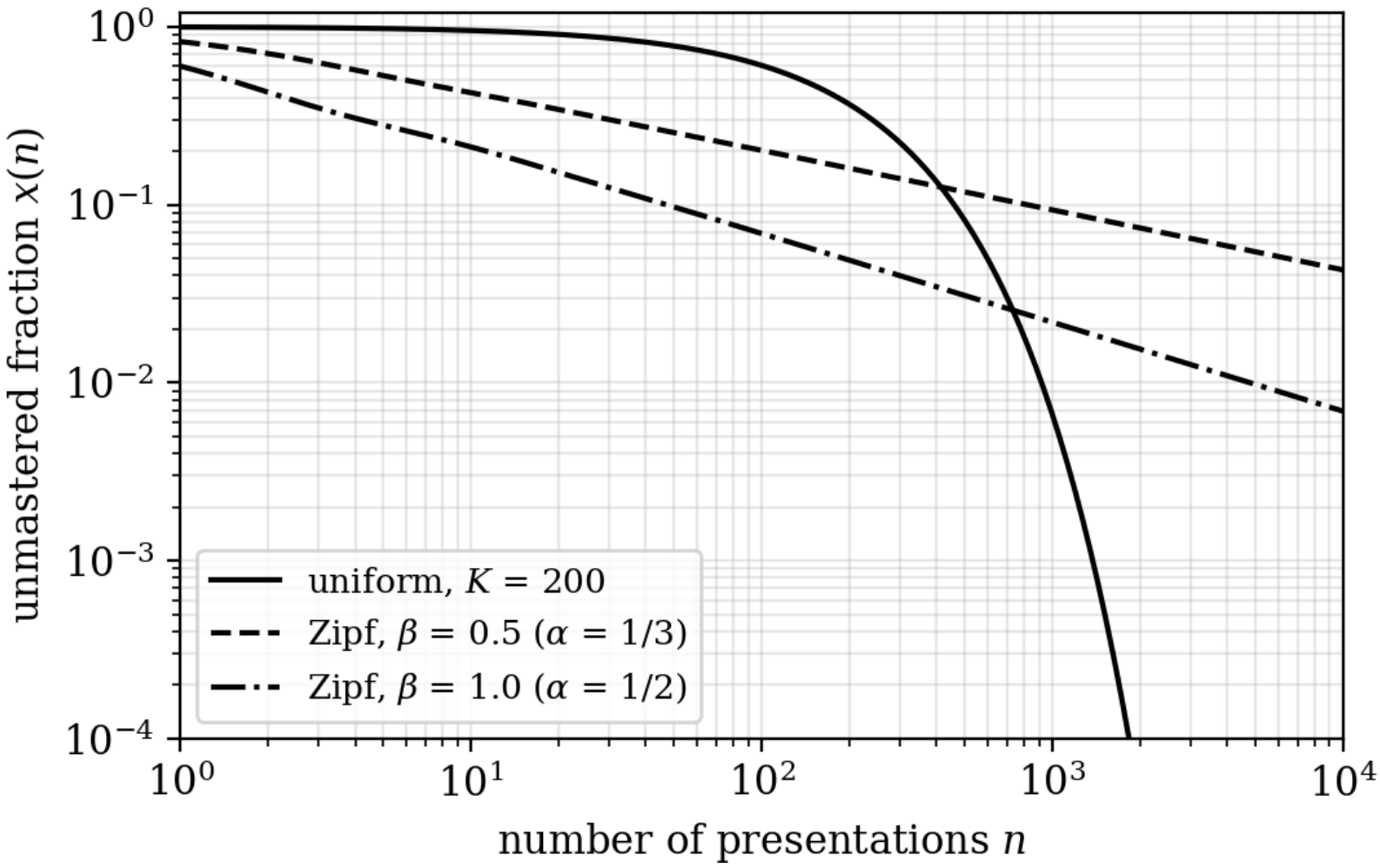

*Fig. 3. Unmastered fraction under uniform and power-law distributions of situation occurrence*

The same result was obtained independently in the theory of artificial neural networks (ANNs) as part of the explanation of scaling laws – under the names "learning curve theory" [33] and "the quantization model of scaling" [42], where the role of the UF values is played by discrete "*quanta*" of skill, mastered in decreasing order of their frequency in the training corpus. The coincidence is exact: formula $\alpha = \beta/(1+\beta)$ was obtained in the theory of activity on the material of debugging, assembly and repair work [27] and in deep learning theory on the material of language models, by different means and with a time gap that rules out borrowing.

## 6. Iteration-Wise Learning Curves

**Problem statement** (simplified)**.** Given: a fixed training set $S$ of size $D$ and a loss function $F(\theta)$ viewed as a function of the parameter vector; an iterative algorithm for its minimization (gradient descent[6]) with step size $\eta_t$. Required: find the form of the dependence of the mismatch $x_t = F(\theta_t) - F^*$ on the computational iteration number $t$, establish which properties of the minimized function (strong convexity, smoothness, the spectrum of the Hessian, gradient-estimation noise) determine the shape of the curve, and pose the problem of choosing $\{\eta_t\}$. Note the difference from the previous section: here the sample is not replenished, and what decreases is the incompleteness of the computation rather than ignorance of the environment.

Let training consist in minimizing a loss function $F(\theta)$ by the gradient method with step size $\eta$. If $F$ is $\mu$-strongly convex and $\Lambda$-smooth, then [46]

(21) $x_t \leq (1 - \eta\mu)^t\, x_0 \approx x_0\, e^{-\eta\mu t}$,

where $x_t = F(\theta_t) - F^*$. This coincides with expression (1) with rate of learning $\gamma = \eta\, \mu$: **a constant per-iteration "contraction" coefficient generates an exponential learning curve** – the mechanism described in the cybernetic models of the mid-twentieth century in the language of the first-order lag element and radioactive decay [49]. Relaxing the assumptions immediately turns the exponential into a power law: for convex smooth functions $x_t = O(1/t)$, for nonsmooth ones $O\left(1/\sqrt{t}\right)$, for the accelerated method $O(1/t^2)$ [46, 16].

A finer description is given by spectral analysis. In the quadratic approximation of the mismatch with Hessian $G$ having eigenvalues $\lambda_i$, the error splits into independent modes:

(22) $x_t = \sum_i c_i\, (1 - \eta\lambda_i)^{2t} \approx \sum_i c_i\, e^{-2\eta\lambda_i t}$.

Expression (22) is a *mixture of exponential learning curves* in which the role of the individual rates of learning is played by the eigenvalues. From this follows the spectral analogue of the classical result that averaging exponential curves over gamma-distributed rates yields a power-law curve [27, 2, 44]. The hypothesis of gamma-distributed individual rates of learning also has direct empirical

[6] *Gradient descent is an iterative minimization method that at each step shifts the parameter vector in the direction opposite to the gradient of the function being minimized; the size of the shift is proportional to the parameter $\eta$, called the step size (learning rate). Stochastic gradient descent computes the gradient on a random part of the sample rather than on the whole of it, which injects noise with variance $\sigma^2$ into the process; one complete pass of the algorithm over the whole sample is called an epoch.*

confirmation – both on data about the formation of cognitive skills in large samples [67] and on classical production data on the execution times of assembly operations [27].

**Proposition 2.** *Let the distribution of the "energy" of the initial error over the spectrum of the mismatch be given by the density of a gamma distribution with shape parameter* α *and scale* b*. Then the optimization curve* (22) *is the power-law curve*

(23) $x_t = x_0\,(1 + 2\eta\theta t)^{-\alpha}$.

**Proof** follows directly from the form of the moment generating function of the gamma distribution, $\mathbb{E}\, e^{-\lambda u} = (1 + bu)^{-\alpha}$, for $u = 2\eta t$. •

If instead the spectrum is a power law, $\lambda_i \asymp \Lambda i^{-(1+\beta)}$, and the mode contributions are proportional to the eigenvalues, then the computations in the proof of Proposition 1 apply verbatim with *D* replaced by $2\eta t$, and we obtain $x_t \approx t^{-\beta/(1+\beta)}$.

Thus **three different "causes" – heterogeneity of the learners, heterogeneity of situation occurrence, and heterogeneity of the spectrum – yield one and the same shape of the learning curve with one and the same exponent.** The family of power-law curves (23) for various values of $\alpha$ is shown in Fig. 4 alongside the exponential; one can see that the difference shows up only at large times and over a wide dynamic range of measurements, so that telling the mechanisms apart from empirical data is practically impossible.

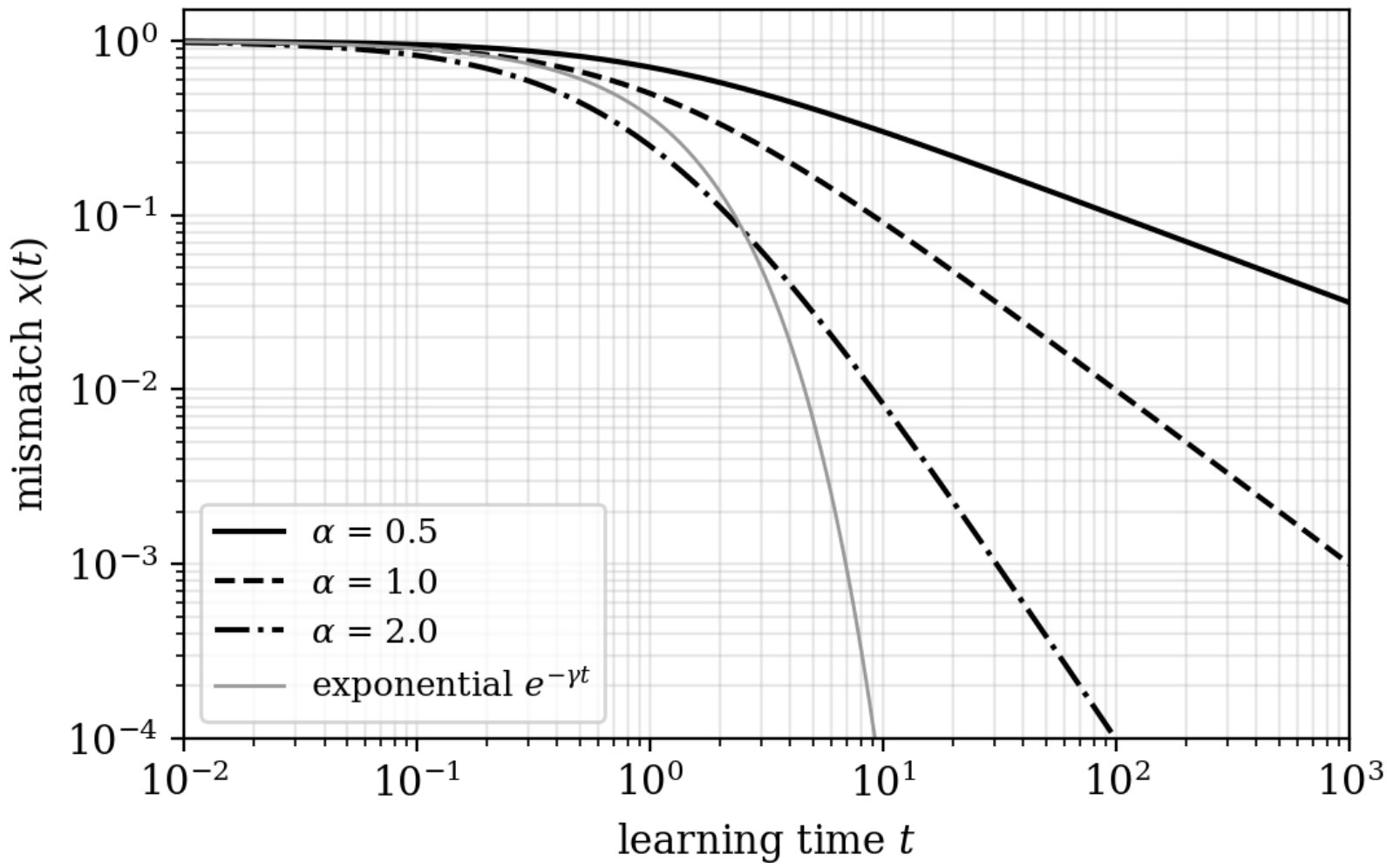

*Fig. 4. A mixture of exponential curves over a gamma-distributed spectrum*

For example, for optimization by stochastic gradient descent with a constant step size and noise variance $\sigma^2$, one has [16]

(24) $\mathbb{E}\, x_t = (1 - \eta\mu)^t x_0 + \frac{\eta\sigma^2}{2\mu}[1 - (1 - \eta\mu)^t]$,

that is, the learning curve has the form (1) with the nonzero asymptote $x_\infty = \eta\sigma^2/(2\mu)$, called in machine learning the *noise floor*. In this case the asymptote is proportional to the step size: in the terms of Section 2, the "physiological limit of learning" is here a controllable parameter, as in model O4, where it turned out to equal $w/(w + u)$.

The classical convergence conditions of *stochastic approximation* for linear problems [73, 57]

(25) $\sum_t \eta_t = \infty, \qquad \sum_t \eta_t^2 < \infty$

lead to the following proposition.

**Proposition 3.** *For a constant step size $\eta_t \equiv \eta$ the learning curve is exponential and has the nonzero asymptote $\eta\sigma^2/(2\mu)$; for $\eta_t = a/t$ with $a > 1/(2\mu)$ the learning curve is hyperbolic, $\mathbb{E}\, x_t = O(1/t)$, and the asymptote is zero.*

A decreasing step-size sequence $\eta_t$ corresponds formally to the "fatigue curve" $w(t)$ in models of learning with fatigue and rest: in both cases the intensity of mastering decreases with time. The two cases differ, however, in the sign of the evaluation. In fatigue models the decrease of $w(t)$ is an undesirable effect to be compensated by rest; in stochastic approximation the decrease of $\eta_t$ is a deliberately chosen control without which convergence to zero error is impossible. Hence the problem of choosing **an optimal schedule of learning intensity** (in machine learning, the *learning rate schedule*) is the classical problem of optimal iterative learning [51], freed from the assumption that the learning level depends only on the accumulated amount of work.

### *7. Scaling Laws*

**Problem statement.** Assume that the three resources – model size $M$, sample size $D$, and compute budget $R$ – are linked by the "semi-empirical" relation $R \approx 6\ M\ D$ (the factor six comes from an estimate of the number of computational operations performed by one model element per input combination), and that the form of the dependence of the loss $F(M, D)$ on the first two has been established empirically. Required: allocate a given budget $R$ between $M$ and $D$ so as to minimize the loss, and find the dependence of the attainable loss on the budget.

The empirical observation[7] [30, 34] is that, for a wide class of ANN architectures, the test-set error is described by power-law dependences in each of the three resources – model size, sample size, and amount of computation – with exponents stable over several orders of magnitude of the argument. The parameterization used is [31]

(26) $F(M, D) = x_\infty + \frac{A}{M^\alpha} + \frac{B}{D^\beta},$

where $x_\infty$ is the irreducible component, playing the same role as the learning limit in expression (1). This gives rise to an optimal resource-allocation problem fully analogous to the classical problems of optimal iterative learning.

[7] *In part of the professional community these results are met with doubt and with suspicions of bias and of examples being cherry-picked to support the conclusions.*

**Proposition 4.** *Let the loss function be given by expression* (26)*, and let the budget* $R = 6MD$ *be fixed. Then the optimal values equal*

(27) $M^* = \left(\frac{\alpha A}{\beta B}\right)^{1/(\alpha+\beta)} \left(\frac{R}{6}\right)^{\beta/(\alpha+\beta)}, D^* = \left(\frac{\beta B}{\alpha A}\right)^{1/(\alpha+\beta)} \left(\frac{R}{6}\right)^{\alpha/(\alpha+\beta)},$

*and the attainable value of the loss function is*

(28) $F^*(R) = x_\infty + \Theta\, R^{-\alpha\beta/(\alpha+\beta)}$.

**Proof.** Write $c = R/6$ and minimize $AM^{-\alpha} + BD^{-\beta}$ subject to the constraint $MD = c$. The first-order conditions of the Lagrange multiplier method give $\alpha AM^{-\alpha} = \beta BD^{-\beta}$, that is, at the optimum the two terms stand in a constant ratio. Substituting $D = c/M$, we obtain $M^{\alpha+\beta} = \frac{\alpha A}{\beta B} c^\beta$, which yields (27); substitution into (26) gives (28). •

The empirical estimates of [31] give $\alpha \approx \beta \approx 0{,}34$, that is, $M^* \propto R^{1/2}$ and $D^* \propto R^{1/2}$.

Four complementary mechanisms are known to generate dependences of type (26).

1. *Frequency-based*: a power-law occurrence distribution of the "units of the task" [33, 42]. This is Proposition 1 of the present paper.
2. *Spectral*: a power-law spectrum of the kernel generated by the network and a power-law expansion of the target function in its eigenfunctions [7, 15, 21]. This is Proposition 2 of the present paper.
3. *Geometric*: the exponent is determined by the intrinsic dimension $d$ of the data manifold, for example $\alpha \approx 4/d$ [62]. This is expression (18) written in terms of the effective rather than the nominal dimension.
4. *Compositional*: a hierarchical generative grammar with a power-law distribution of productions [20]. This is a formalization of the thesis that an action is a composition of operations.

Thus both lines of explanation of power-law learning curves, developed independently in the theory of activity on the material of production tasks [27], have been reproduced in deep learning theory, with coinciding formulas for the exponent. This is an argument for the generality of the regularities under consideration. On the other hand, although the regularities are shared, the mechanisms of their realization, their interpretation, and their control in machine learning differ substantially from those in human activity.

## *8. Learning Curves in Reinforcement Learning*

**Problem statement.** Given: an agent interacting with an environment; at each step $t$ it observes the environment's state, chooses an action, and receives a scalar reinforcement $r_t$ and a new state; the required outcome is not presented explicitly. The quantity $V_t$ is the current value estimate, that is, of the expected total reinforcement. Required: find the shape of the learning curve in the number of value-update iterations and in the number of real trials, that is, interactions with the environment, and compare it with the curves of the classical models of animal learning.

The classical models of animal learning – the stochastic model of R. Bush and F. Mosteller [18] and C. Hull's Postulate IV [32] – are the direct predecessors of the Rescorla–Wagner rule [56], which in turn is a special case of the temporal-difference method [68]:

(29) $V_{t+1} = V_t + \eta\,(r_t - V_t)$.

For constant $\eta$ and stationary reinforcement, the expectation of the mismatch satisfies the relation

(30) $\mathbb{E}\,(V_\infty - V_t) = (1-\eta)^t\,(V_\infty - V_0)$,

that is, the learning curve is exponential with rate $\gamma = -\ln(1-\eta) \approx \eta$. This is the curve obtained for the animal in the maze in the 1950s.

In the classical setting, reinforcement learning reduces to finding the fixed point of the Bellman operator, which is a contraction with coefficient $\chi \in (0; 1)$:

(31) $\parallel V_t - V^* \parallel \le \chi^t \ \parallel V_0 - V^* \parallel$.

We again obtain an exponential learning curve generated by the constancy of the contraction coefficient. At the same time the *sample complexity*, that is, the number of real interactions with the environment required to reach an $\varepsilon$-optimal policy, is of order $\tilde{O}(|\Omega||Y|\varepsilon^{-2}(1-\chi)^{-3})$ [6], so the learning curve in the number of "trials" is a power law with exponent $\alpha = 1/2$ (for nonstationary and/or nonlinear reinforcement learning mechanisms this condition may fail to hold; further study is needed here).

The combination of an exponential curve in computational iterations and a power-law curve in the number of examples is the characteristic signature of a *two-level learning scheme*, in which a fast inner loop (recomputing the estimates) is nested inside a slow outer one (accumulating the experience of interaction with the environment). This conclusion also has a converse reading: observing both an exponential and a power-law curve in a learning system, over different arguments, one should look for a two-level structure of the process rather than for a contradiction.

## *9. Nonmonotone Learning Curves*

**Problem statement.** Learning curves observed in machine learning systematically violate W. Ashby's monotonicity principle ("never lose what has been gained")[8] as an empirical regularity of the functioning of biological systems [3, 4]. Required: first, specify under what minimal assumptions the general model of experience formation of Section 3 can generate nonmonotone curves, and, second, pose the practical problem arising from this – extrapolating the learning curve from its initial segment and deciding when to stop training (which is hardly ever done in machine learning).

Almost all the curves considered above are monotone; in machine learning such curves are called "*well behaved*". At least four classes of "nonmonotonicity" phenomena have been noted.

– *Double descent* [8, 45]: the test error is nonmonotone in the number of parameters and in the number of epochs, reaching a local maximum at the interpolation threshold $M \approx D$. Nonmonotonicity is possible in the sample size as well [40].

– *Delayed generalization* (*grokking*) [55]: a long plateau followed by an abrupt transition to nearly zero error. Formally this is a curve with an intermediate plateau (Section 2), but in practice the plateau's length can exceed the duration of the first phase by two to three orders of magnitude.

– *Emergent effects* [79]: the abrupt appearance of new abilities once a scale threshold is crossed. In [60] it was shown for a particular model that the jump is to a considerable extent an artifact of the threshold (dichotomous) metric: switching to a continuous metric makes the curve smooth again. This is the remark of Section 2 that with dichotomous scales the appearance of step-like and logistic curves is unnatural and can be removed by choosing an appropriate scale and units of measurement.

[8] *The interpolation threshold is the regime in which the number of tunable parameters suffices to reproduce all training examples exactly, that is, the empirical risk vanishes.*

– *Broken power laws* [19]: a change of the exponent when a characteristic scale is crossed; the direct analogue of the piecewise power-law individual learning curves noted in psychology.

Nonmonotonicity has a direct consequence for the general model of experience formation.

**Proposition 5.** *If in the general model of experience formation (Section 3) the probability of mastering is* $w(t) \geq 0$ *and the probability of forgetting is* $u \equiv 0$*, then the learning curve is nondecreasing for any law of variation of* $w(\cdot)$.

**Proof.** From the difference equation (4) with $U \equiv 0$ it follows that $q(t+1) - q(t) = w(t)\big(1 - q(t)\big) \geq 0$ for $q(t) \leq 1$. •

Consequently, **any empirically observed nonmonotone learning curve requires introducing either nonzero forgetting $u > 0$ or nonstationarity of the UF distribution.** The first corresponds to models with forgetting and fatigue; the second takes the phenomenon outside iterative learning in the sense of the definition of Section 2, since the constancy of external conditions is violated. In other models one could, presumably, find other "explanations". Thus the nonmonotone curves of machine learning are consistent with the construction presented here and indicate which of its branches should be applied.

Of practical importance is the problem of *extrapolating* the learning curve: predicting the asymptote from an initial segment and deciding when to stop training. Standard theoretical approaches use mixtures of parametric curve families [23], Bayesian neural networks [37], and Gaussian processes with a special kernel reflecting exponential decay [69]. This is the problem of identifying a model of iterative learning from an incomplete realization.

## *10. Partitioning the Situation Space and Typical Solutions*

So far the distribution $\{p_k\}$ of UF value occurrence has been considered given. In problems of designing technologies and training systems, however, it is a matter of choice: the designer decides into how many situations to partition the space of environment states and how coarse to make these situations. The corresponding problem has a normative solution.

*10.1. UNIFORM PARTITION*

**Problem statement.** Given: a continuous or very large set of situations $\Omega$ and the number $K$ of situation classes the system distinguishes, that is, of partition cells[9]. Required: choose a partition of the set $\Omega$ into $K$ subsets, that is, a set of probabilities $\{p_k\}$, maximizing the expected learning level by a given time $t$ and minimizing the expected error of the decisions made, and also study the dependence of the error on the number of cells $K$. This is the question of how finely the system should distinguish situations.

Consider an agent distinguishing $K$ UF values realized with probabilities $p_k$ and deeming a component of experience mastered at the very first realization of the corresponding value. The learning level equals

(32) $L_t = 1 - \sum_{k=1}^{K} p_k \ (1 - p_k)^t,$

that is, it coincides with expressions (6) and (19). Thus the machinery of experience models carries over to machine learning unchanged.

Let the efficiency of the agent's decisions be described by a function $\varphi(y, \omega)$, uniformly $l$-Lipschitz in its first argument, and let the set of states of nature be partitioned into $K$ subsets, each associated with one *typical solution*: at the first realization of a state from a given subset the solution is found and memorized; at a repeat realization the ready-made one is applied. Then the expected error of the decisions made is estimated by the expression

(33) $F(\{p_k\}, t) = l \ \sum_{k=1}^{K} p_k^2 \ (1 - p_k)^t,$

and in the case of the uniform distribution by

(34) $F_0(K, t) = \frac{l}{K}\left(1 - \frac{1}{K}\right)^t.$

The first factor $p_k$ in (33) corresponds to the probability of landing in the $k$-th cell, the second to the cell's characteristic size, and $(1 - p_k)^t$ to the probability that the cell has not yet been encountered.

[9] *Quantization is the replacement of the value of a continuous quantity by the nearest element of a finite set, called in machine learning a codebook. For texts, roughly the same role is played by tokenization – splitting the text into elements of a finite vocabulary.*

The first two factors give the classical error bound for the piecewise-constant approximation of a Lipschitz function; the third reflects the incompleteness of experience.

Let a threshold $\rho$ for distinguishing UF states be given.

**Proposition 6** [10]. *For any threshold $\rho$ of distinguishing states of nature (UF) there exists a moment of time starting from which the unique solution of the problem of minimizing the expected error* (33)*, as well as of the problem of maximizing the expected learning level* (32)*, is the uniform partition, $p_k = 1/K$.*

Proposition 6 also survives when one accounts for the costs $C_k$ of finding a solution at the first realization of a situation, and for the gains $H_k$ from applying a ready-made solution at repeat realizations [10].

In machine learning terms, this is a statement about the choice of discretization of the input space, and it admits at least four readings.

1. *Quantization and tokenization.* Partitioning the feature space into $K$ cells is the construction of a vocabulary or a codebook [26]. Proposition 6 says that the cells should be made equiprobable rather than equal in volume, i.e., quantile quantization is optimal (algorithms that grow the vocabulary by merging frequent pairs thereby equalize element frequencies).
2. *Codebook collapse and expert balancing.* A known problem in training discrete autoencoders [52] is that only a small fraction of the code vectors gets used, that is, the distribution $\{p_k\}$ degenerates. The remedies – resetting unused codes and entropy regularization – are frequency-equalization techniques. The same logic underlies the penalty for uneven expert load [25, 63]. Proposition 6 gives these techniques a normative justification: uniformity here is an optimum in its own right.
3. *Consistency of the entropy and the "learning" criteria.* The uniform distribution maximizes entropy [10] and simultaneously maximizes the expected learning level. Two generally different criteria have a common optimum here, which indirectly justifies the widespread use of entropy as a surrogate for representation quality.
4. *The optimal number of cells and double descent.* Analysis of expression (34) yields the following result.

**Proposition 7** [10]. *For any* $t \geq 0$ *there exists a unique worst (maximizing the expected error) number of cells* $K^*(t) = t + 1$.

**Proof.** Setting $u = 1/K$ and treating $F_0 = l\,u(1-u)^t$ as a function of a continuous argument, we obtain $\partial F_0/\partial u = l(1-u)^{t-1}[(1-u) - tu]$, whence the unique stationary point $u = 1/(t+1)$, which delivers the maximum. •

In other words, the error is nonmonotone in the model's "complexity" and reaches its maximum when the number of distinguished situations is roughly equal to the number of trials conducted (Fig. 5). This is a structural analogue of *double descent* [8, 45], where the error peak falls at the interpolation threshold, that is, at the equality of the number of parameters and the sample size.

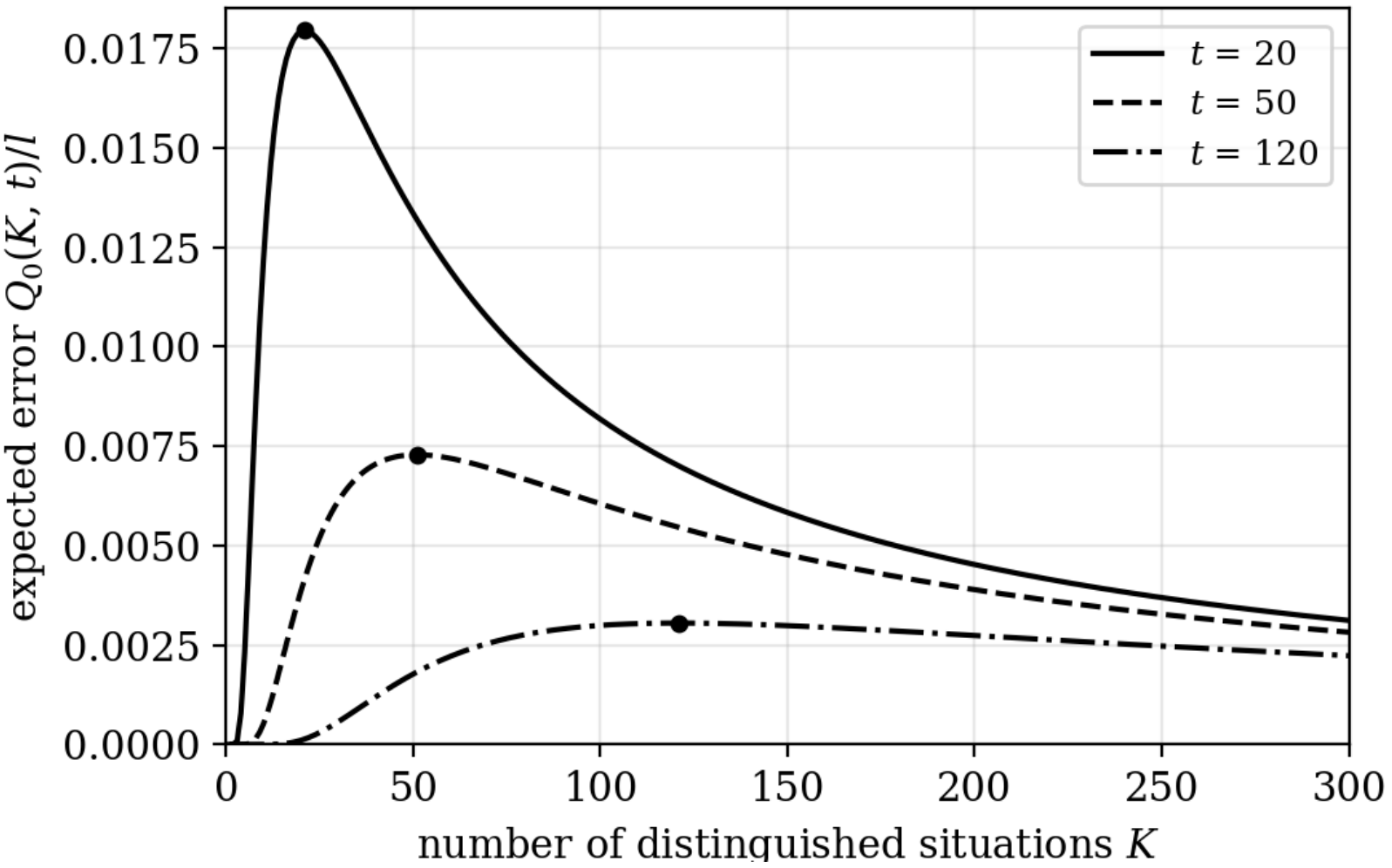

*Fig. 5. Nonmonotonicity of the expected error in the number of distinguished situations*

The limits of the analogy should be stated. The mechanisms differ: in double-descent theory the peak is generated by the growth of the variance of the interpolating estimate, whereas in model (34) it is generated by the fact that at $K \approx t$ a substantial fraction of the cells has not yet been encountered even once, while the cells themselves are already fine enough for each one's contribution to be noticeable. What is common is the shape of the dependence: the theory predicts non-monotonicity of the error in the number of distinguished situations, with a maximum at the point where the model's complexity matches the amount of experience. The value of Proposition 7 is that it gives this effect in closed form and with the exact location of the maximum, which double-descent theory does not yet provide.

### *10.2. TYPICAL SOLUTIONS*

**Problem statement.** Given: a partition of the situation set into $K$ cells; each cell stores one *typical solution*, found at the first entry into it and applied at all subsequent ones. Required: establish which constructions of machine learning correspond to typical solutions, and indicate the limits of this correspondence[10].

The construction of typical solutions [77, 50], widespread in situational and adaptive control [54], has analogies in machine learning along the following lines.

– *Prototype classifiers*: the $K$-means method, learning vector quantization [38], prototypical networks in few-shot learning [66]. A class prototype is the typical solution for its cell.

– *Codebooks*: discrete autoencoders [52], as well as quantization of network weights and activations. The codebook size is the parameter $K$ of problem (33).

– *Mixtures of experts* [25, 63]: the router partitions the input space, and each expert implements the typical solution of its region. The problem of choosing the number of experts is the problem of choosing $K$, and the penalty for uneven load is an implementation of Proposition 6.

– *Piecewise-constant models*: decision trees and forests, where a leaf is a cell and the value in the leaf is the typical solution. Bound (33) with $t \to \infty$ turns into the standard bound $l/K$ for the uniform partition.

---

[10] *An autoencoder is an encoder–decoder pair tuned to reconstruct a situation from its compressed description; in a discrete autoencoder, the compressed description is the index of a codebook element. A mixture of experts is a technology composed of several submodels ("experts") and a router that decides, for each situation, which expert to pass it to. A decision tree is a technology that successively splits the set of situations by tests; its leaves store the answers.*

– *Caching and pattern lookup*: the rule "if the state has recurred, apply the previously found solution" is a cache, and $L_t$ is its hit rate. Language models with external memory [35] implement the same scheme (of course, LLMs also use a subtler approach, namely semantic matching, which is not an analogue of "the state has recurred"), and the empirical hit-rate curves have the form (32).

One limitation, however, must be stated whenever results are transferred. Classical quantization theory [26] optimizes the mean squared reconstruction error and, at high resolution, yields nonuniform cells whose density is proportional to the cube root of the distribution density. Criterion (33) is different: it weights the error by the probability that the situation has not yet been encountered, that is, it optimizes experience coverage, with reconstruction accuracy left aside. Hence the probability uniformity resulting from Proposition 6 is compatible with the nonuniformity of classical quantizers, since the two answer different questions. Moreover, classical quantization deals with linear problems, whereas almost all machine learning problems are nonlinear, so the conclusions and mechanisms of the classical theory work, in practice, in a very small domain. In practice, vocabularies and codebooks intended for mastering new situations should be balanced by frequencies, and those intended for accurate reconstruction by the model's errors.

## *11. The Problem of Choosing the Training Distribution*

**Problem statement.** Given: the number $K$ of UF values; the *deployment* distribution $\{p_k\}$ with which the trained system will encounter situations in operation (fixed externally and not subject to change); the training budget – the number $D$ of examples presented. Required: choose the *training* distribution $\{\pi_k\}$ with which examples are presented during training so as to maximize the expected learning level measured under deployment conditions[11]. Unlike the problem of Section 10, here training and deployment take place under different conditions, and the controlled variable is the composition of the training set rather than the partition of the situation set.

*[11] A language model is a technology that estimates the probability of the next element of a text given the preceding ones. The training corpus of such a model is assembled from sources of different nature (books, program code, reference works) and semantics, and the shares of these sources in the corpus are what is chosen in the problem under consideration.*

Proposition 6 answers the question of how to partition the situation space when training proceeds under the same conditions in which the system will then operate. In practice this assumption can easily be violated: the training set is put together by the developer, while the deployment conditions are fixed externally. Let us pose the corresponding problem.

Let the UF take $K$ values, and let the *deployment* distribution $\{p_k\}$ be known and not subject to change: it is with these frequencies that the trained system will encounter situations in operation. Training consists in presenting $D$ independent examples generated by a *training* distribution $\{\pi_k\}$ that we are free to choose. Within model O1, a component of experience is mastered at the very first realization of the corresponding UF value, so after $D$ presentations it has been mastered with probability $1-(1-\pi_k)^D$. The expected learning level, measured under deployment conditions, equals

(35) $L(D \mid \pi) = \sum_{k=1}^{K} p_k \left[1-(1-\pi_k)^D\right] = 1 - \sum_{k=1}^{K} p_k \ (1-\pi_k)^D$.

The problem is to maximize (35) over $\pi$ on the simplex $\sum_k \pi_k = 1$, $\pi_k \geq 0$. This is the question of which distribution to train on a system that will have to operate on a different one.

**Proposition 8.** *The problem of maximizing* (35) *has a unique solution. If it is interior, it is determined by the condition*

(36) $p_k\,(1-\pi_k)^{D-1} = \text{const}, \qquad k = 1, \dots, K,$

*and has the form*

(37) $\pi_k^* = 1 - \frac{(K-1)\,p_k^{-\alpha}}{\sum_{j=1}^{K} p_j^{-\alpha}}, \qquad \alpha = \frac{1}{D-1},$

*and the attained value of the criterion equals*

(38) $L^*(D) = 1 - \frac{(K-1)^D}{J^{D-1}}, \qquad J = \sum_{j=1}^{K} p_j^{-\alpha}.$

**Proof.** The function $(1-\pi_k)^D$ is convex in $\pi_k$ for $D \geq 1$, so the function to be minimized, $\sum_k p_k\,(1-\pi_k)^D$, is convex, and the feasible set is a simplex; existence and uniqueness of the solution follow, and the Karush–Kuhn–Tucker conditions are sufficient. Setting the derivatives of the Lagrangian to zero, we obtain (36). Hence $1-\pi_k = c\,p_k^{-\alpha}$; substitution into the normalization condition gives $c =$ $(K-1)/J$ and expression (37). Substituting (37) into (35) and using the fact that $\alpha D - 1 = \alpha$, we obtain $L^* = 1 - c^D \sum_k p_k^{1-\alpha D} = 1 - c^D J$, which coincides with (38). •

**Proposition 9.** *The optimal training distribution* (37) *has the following properties.*

1. *Monotonicity*: $\pi_k^*$ strictly increases in $p_k$, that is, the order of the frequencies is preserved – situations more frequent in deployment receive a larger share of training (though there is a risk of obtaining a degenerate model with an uncontrolled type-II error).
2. *Limiting behavior*: as $D \to \infty$ we have $\alpha \to 0$ and $\pi_k^* \to 1/K$ for all $k$, that is, the optimal training distribution tends to the uniform one regardless of the deployment distribution. Thus Proposition 6 is the limiting case of Proposition 8 for an unbounded training budget.
3. *Small budget*: for small $D$ the optimal training distribution is concentrated on frequent situations more strongly than the deployment one. There exists a budget value at which the training and deployment distributions are closest; below that value the advantage goes to concentration, above it to equalization.
4. *Cutting off rare situations*: if for some $k$ the right-hand side of (37) is negative, the solution is a boundary one, $\pi_k^* = 0$, and the corresponding situation is excluded from training entirely. The optimal support is found as follows: the values $p_k$ are ordered decreasingly and one takes the largest $m$ for which all $\pi_k^*$, computed by (37) with $K$ replaced by $m$ and summation over the first $m$ values, are strictly positive.

In substance the last property means the following: **with a limited training budget, rare situations should be neglected entirely; allotting them a small share of examples is worse. This theoretical conclusion contradicts the widespread practice of "giving at least a few examples of every class", and the explanation is that, for linear algorithms,** an example spent on a situation with a negligible deployment probability is taken away from a situation that will occur almost surely.

**Example.** Let $K = 4$ and $p = (0{,}50;\ 0{,}30;\ 0{,}15;\ 0{,}05)$. The optimal training distributions are given in Table 3; zeros mean the exclusion of a situation from training. The dynamics of the optimal shares as the budget grows is shown in Fig. 6.

*Table 3. Optimal training distribution for* $p = (0{,}50;\ 0{,}30;\ 0{,}15;\ 0{,}05)$

| $D$ | $\pi_1^*$ | $\pi_2^*$ | $\pi_3^*$ | $\pi_4^*$ | $L^*(D)$ |
|---|---|---|---|---|---|
| 2 | 0.625 | 0.375 | 0 | 0 | 0.613 |
| 5 | 0.427 | 0.348 | 0.225 | 0 | 0.842 |
| 10 | 0.332 | 0.293 | 0.237 | 0.138 | 0.960 |
| 20 | 0.289 | 0.270 | 0.243 | 0.198 | 0.998 |
| 100 | 0.258 | 0.254 | 0.249 | 0.240 | 1.000 |

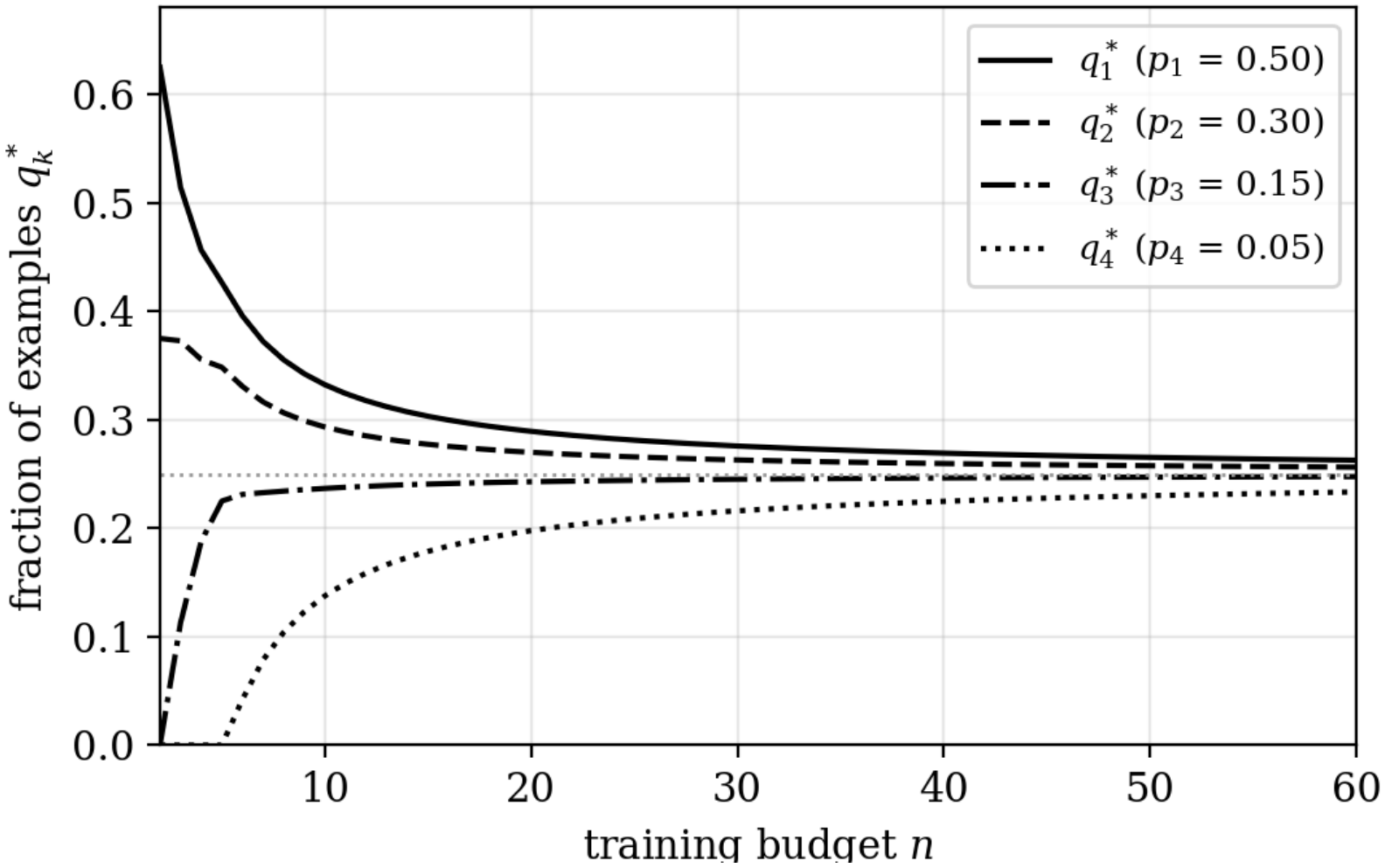

*Fig. 6. Optimal training distribution as a function of the training budget*

For $D = 2$ training is concentrated on the two most frequent situations; for $D = 20$ the distribution is already “close to uniform”. Let us compare three strategies (Table 4): proportional ($\pi = p$, copying the deployment distribution), uniform ($\pi_k = 1/K$), and optimal.

*Table 4. Comparison of three training strategies (cells give the expected learning level for a given D)*

| $D$ | Proportional ($\pi \equiv p$) | Uniform ($\pi_k = 1 / K$) | Optimal (see (37)) |
|---|---|---|---|
| 2 | 0.575 | 0.438 | 0.613 |
| 5 | 0.829 | 0.763 | 0.842 |
| 20 | 0.976 | 0.997 | 0.998 |

**The preference order of the two simple strategies reverses as the training budget grows**: for short training it is better to copy the deployment distribution (see Table 4: at $D = 2$, 0.575 > 0.438), for long training to equalize it (see Table 4: at $D = 20$, 0.997 > 0.976).

The result obtained relates directly to several strands of machine learning.

– *Distribution shift and reweighting*. The classical approach to covariate shift [64] prescribes weighting with weights $p_k/\pi_k$. Proposition 8 answers the prior question of what $\pi$ itself should be.

– *Class balancing under long tails*. For example, in [22] it was established empirically that the optimal reweighting lies between no balancing and full balancing, shifting toward the latter as the data volume grows (there are, admittedly, other approaches as well). Formula (37) gives this a quantitative explanation; moreover, the "effective number of samples" $(1 - \xi^D)/(1 - \xi)$ used in that work coincides structurally with the factor $1 - (1 - \pi_k)^D$ in (35).

– *Machine teaching*. Problem (35) is the simplest inverse of the usual problem, in which a model is fitted to the data; here the data are fitted to the goal [84]. Unlike the general setting, here the solution is written out explicitly.

We note natural generalizations that preserve the proof technique of Proposition 8: unequal costs of presenting examples of different types, when the constraint takes the form $\sum_k c_k \pi_k \leq C$; the need for $\nu$-fold repetition for mastering, when the factor $1 - (1 - \pi_k)^D$ is replaced by a tail of the binomial distribution; nonzero forgetting $u > 0$, under which rare situations require regular repetition and cutting off ceases to be optimal.

## *12. Correspondence Between the Concepts of the General Model of Activity and of Machine Learning*

The correspondence between the concepts of the general model of activity and those of machine learning is given in Table 5.

*Table 5. Correspondence between the concepts of the general model of activity and of machine learning*

| **General model of activity** | **Machine learning** |
|---|---|
| Active element (agent) | Trained model, network, policy |
| Value of the uncertainty factor | Training example, environment state |
| Technology of activity | Hypothesis, parameter vector |
| Component of experience mastered | Example processed correctly |
| Probability of mastering $w(t)$ | Step size, learning rate schedule |
| Probability of forgetting $u(t)$ | Catastrophic forgetting[12], interference |
| Learning level $L(t)$ | One minus normalized error |
| Agent's farsightedness | Planning horizon, training budget |
| Aggregation of components | Compositional structure of the task |
| Optimal allocation of work | Optimal allocation of the budget |
| Typical solution | Prototype, codeword, expert, tree leaf |
| Uniform partition | Quantile quantization, expert balancing |

The row corresponding to forgetting deserves special attention: *catastrophic forgetting* [41] – the loss of mastered abilities when learning a new task – is described by models with a nonzero probability $u$, and the remedies applied against it [36] amount to decreasing $u$ for the components of experience deemed essential. Curriculum learning problems [14, 84] are problems of optimal iterative learning [51] with the assumption dropped that the learning level is independent of the schedule.

---

*[12] Catastrophic forgetting is the loss of previously mastered abilities when learning a new task: tuning the parameters to the new examples destroys the configuration that ensured correct answers on the old ones.*

## *13. Conclusion*

Let us discuss the main results (beyond the general thesis of the unity of learning regularities in natural and artificial systems).

The learning curve in machine learning has three arguments – sample size, number of iterations, and compute budget; mixing these sections up leads to terminological confusion.

Exponential learning curves arise where each “step” removes a fixed share of uncertainty (contraction operators, strongly convex problems, stationary reinforcement, a homogeneous environment); power-law curves where the heavy tail of rare situations is being “worked through” or heterogeneous rates are averaged.

The exponent of the power-law learning curve is expressed through the exponent of the situation-occurrence distribution by the formula $\alpha = \beta/(1+\beta)$ (Proposition 1), obtained independently in the theory of activity and in the theory of scaling laws for ANNs.

A mixture of exponential curves over a gamma-distributed spectrum yields a power-law curve (Proposition 2). The indistinguishability of the three mechanisms from the observed curve has a direct methodological consequence, discussed below.

The learning-rate schedule is formally identical to the fatigue curve, although the fatigue curve is an uncontrollable disturbance whereas the schedule is a control instrument. The corresponding optimization problem is a substantive generalization of the problem of optimal iterative learning.

Nonmonotone curves (double descent, delayed generalization) are compatible with the general model and point to the need to account for forgetting or for nonstationarity of the external conditions (Proposition 5).

Proposition 6 on the uniform partition is a normative justification of a whole range of machine learning effects: quantile quantization, codebook and expert load balancing, entropy regularization. Proposition 7, which gives the worst number of distinguished situations $K^*(t) = t + 1$, describes in closed form an effect structurally analogous to double descent: the error is maximal where the model's complexity matches the amount of experience.

Typical solutions have in machine learning such analogues as prototypes, codewords, experts, decision-tree leaves and cache entries,

and bound (33) is the usual error bound for a piecewise-constant model, supplemented by a mastery factor. At the same time, the experience-coverage criterion and the reconstruction-accuracy criterion lead to different optimal partitions, which must be kept apart when transferring results.

If the distribution of situations in operation is given while the training distribution is a matter of choice, then the optimal training distribution (Propositions 8 and 9) is, for a small training budget, concentrated on frequent situations more strongly than the deployment one, and tends to the uniform one as the budget grows; with a limited budget, rare situations may be neglected.

The main outcome of the work is that both lines of explanation of power-law learning curves, developed in the theory of activity on the material of debugging, assembly and repair work, have been independently reproduced in deep learning theory, with coinciding formulas: the frequency mechanism corresponds to Proposition 1, the spectral one to Proposition 2. The exponential curves of machine learning are generated by the constancy of the contraction coefficient – the mechanism revealed half a century earlier in models of radioactive decay and the first-order lag element. The coincidence of results obtained half a century apart, on incomparable material and by different means, should be regarded as yet another independent confirmation of the generality of the regularities of iterative learning in natural and artificial systems.

A second conclusion concerns the limits of inference from the curve. Three distinct "mechanisms" – heterogeneity of the learners, a power-law spectrum of the mismatch, and a heavy tail of the situation-occurrence distribution – generate power-law curves indistinguishable in shape. Consequently, an observed learning curve does not allow one to judge the structure of the learning system, and attempts to draw conclusions from it about that system's nature and internal mechanisms are methodologically incorrect. This is a direct manifestation of the limits of the *inverse method* of model construction [49]: if different assumptions about the laws of interaction lead to one and the same result, then, lacking additional information, one cannot say unambiguously which of the equivalent models corresponds to the real system. This is a result of the theory rather than a defect of it, and it deserves to be taken seriously by those who try to judge the structure of intelligence, natural or artificial, from the learning curve.

Finally, machine learning offers learning theory what it has lacked over the past century: the possibility of a controlled experiment with a learning system of known structure, in which the structure of the environment, the sample size and the forgetting mechanism can be varied at will. The problems of curriculum construction, of combating catastrophic forgetting, of learning-curve extrapolation and of optimal step-size selection are, respectively, the problems of aggregating the components of activity, of forgetting and fatigue, of identifying learning models, and of optimal iterative learning, posed on new material; their joint study appears to be a promising direction for further work.

Separately, we note directions requiring a departure from the assumptions adopted above: abandoning the binarity of experience components and introducing a hierarchical structure on the set of UF values (which would allow skill transfer, interference and generalization to be described within models of learning to learn); an endogenous description of fatigue; and the transition from a stationary situation-occurrence distribution to a nonstationary one generated by the joint evolution of the learner and the environment.